\documentclass{aa}
\usepackage{graphicx}

\newcommand{\HI}{H{\,\small I}}

\newcommand{\smallHI}{H{\,\scriptsize I}}
\newcommand{\smallOIII}{O{\,\scriptsize III}}
\newcommand{\matHI}{\rm H{\hskip 0.02cm\scriptscriptstyle I}}
\newcommand{\Halpha}{H$\alpha$}

\newcommand{\kms}{km s$^{-1}$}
\newcommand{\atms}{atoms cm$^{-2}$}

\newcommand{\mJybeam}{mJy beam$^{-1}$}
\newcommand{\msun}{{$M_\odot$}}

\newcommand{\tspin}{$T_{\rm spin}$}

\input{psfig}

\begin{document}
\title{Large-scale gas disk around the radio galaxy Coma~A\thanks{Based on
    observations with the Westerbork Synthesis Radio Telescope (WSRT) and the
    Very Large Array (VLA).  }}

\authorrunning{ Morganti et al.}
\titlerunning{\smallHI\ absorption in the radio galaxy Coma~A}

\subtitle{}
   
\author{R. Morganti\inst{1}, T.A. Oosterloo\inst{1}, S.Tinti\inst{2}\thanks{
    Netherlands Foundation for Research in Astronomy summer student 2001},
  C.N. Tadhunter\inst{3}, K.A. Wills\inst{3}, \and G. van Moorsel\inst{4} }

\offprints{R.Morganti}
   
\institute{Netherlands Foundation for Research in Astronomy, PO Box 2, 7990
     AA, Dwingeloo, The Netherlands
    \and         Istituto di Radioastronomia, CNR, via Gobetti 101, I-40129
             Bologna, Italy 
    \and     Dep. Physics and Astronomy,
             University of Sheffield, Sheffield, S7 3RH, UK 
    \and     National Radio Astronomy Observatory, Socorro,
             NM 87801, USA 
             }

\date{Received ...; accepted ...}

\abstract{ We present WSRT and VLA radio observations of the neutral hydrogen
  in the radio galaxy Coma~A. We detect extended \HI\ absorption against both
  radio lobes of Coma~A, at distances of about 30 kpc from the centre.  Coma A is
  the first radio galaxy in which \HI\ is seen in absorption at such large
  distances from the nucleus.  The match between the velocities of the neutral
  hydrogen and those of the extended ionized gas suggests that they are part
  of the same disk-like structure of at least 60 kpc in diameter.  Most
  likely, this gas disk is partly ionised by the bulk motion of the radio
  lobes expanding into it.  The gas mass of this disk is at least $10^9$
  \msun. The relatively regular structure of the gas disk suggests that a
  merger occurred involving at least one large gas-rich galaxy, at least a few
  times $10^8$ yr ago. \keywords{ galaxies: ISM -- galaxies: active -- radio
lines: galaxies
 } } 

\maketitle
%

\section{Gas in and around radio galaxies}

Many early-type galaxies in the nearby universe appear to have experienced a
recent merger/accretion event. In many cases, this has involved at least one
gas-rich companion/galaxy that has brought a significant amount of gas into
the galaxy. Indeed, \HI, FIR and CO observations of elliptical galaxies have
demonstrated that many of these objects have an active and interesting cold
interstellar medium, often qualitatively (and sometime also quantitatively)
similar to that observed in spirals (see Knapp 1999 for a review). Gas is a
key factor in determining the evolution of early-type galaxies (e.g.\ 
Kauffmann 1996).  Among early-type galaxies, gas-rich systems may represent an
important phase in the evolution that many elliptical galaxies go through and
they  give  important information on the formation and evolution of these
systems.

Apart from being important for the overall evolution of early-type galaxies,
there is also compelling morphological and kinematical evidence that the
activity in powerful radio galaxies is triggered by galaxy mergers and
interactions (e.g.\ Heckman et al.\ 1986, Smith \& Heckman 1989, De Koff et
al.\ 1996, Verdoes Klein et al.\ 1999, Capetti et al.\ 2000; see e.g.\ Wilson
1996 for a review).  This is also supported by the theoretical results
(Kauffmann \& Haehnelt 2000) that suggest that the evolution of supermassive
black holes is strongly linked to the hierarchical build-up of galaxies.
Although these processes are likely to be more efficient and frequent at high
redshifts, they are also observed in relatively ``nearby'' radio galaxies.

\begin{figure*}
\centerline{\psfig{figure=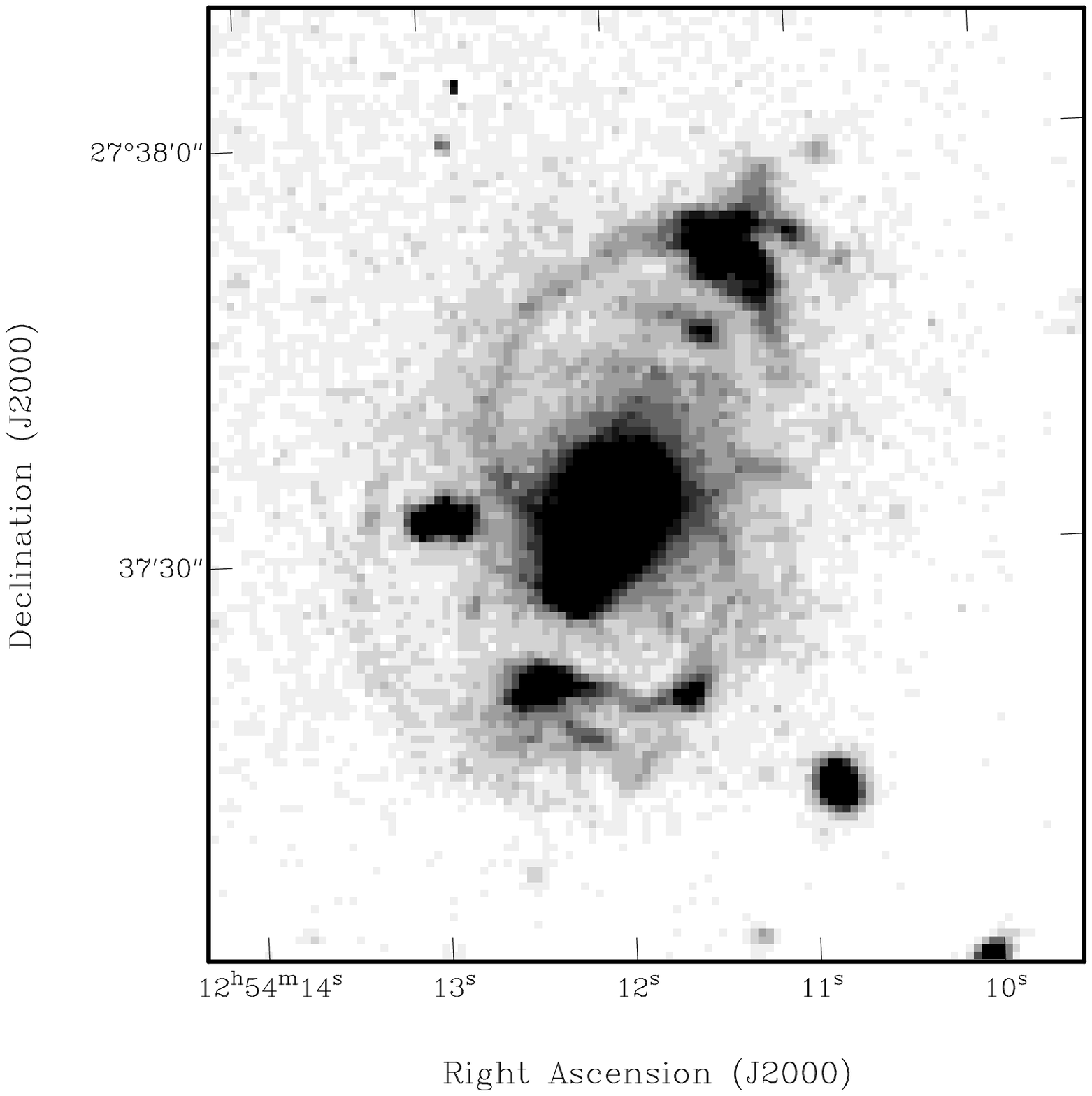,width=8cm}\psfig{figure=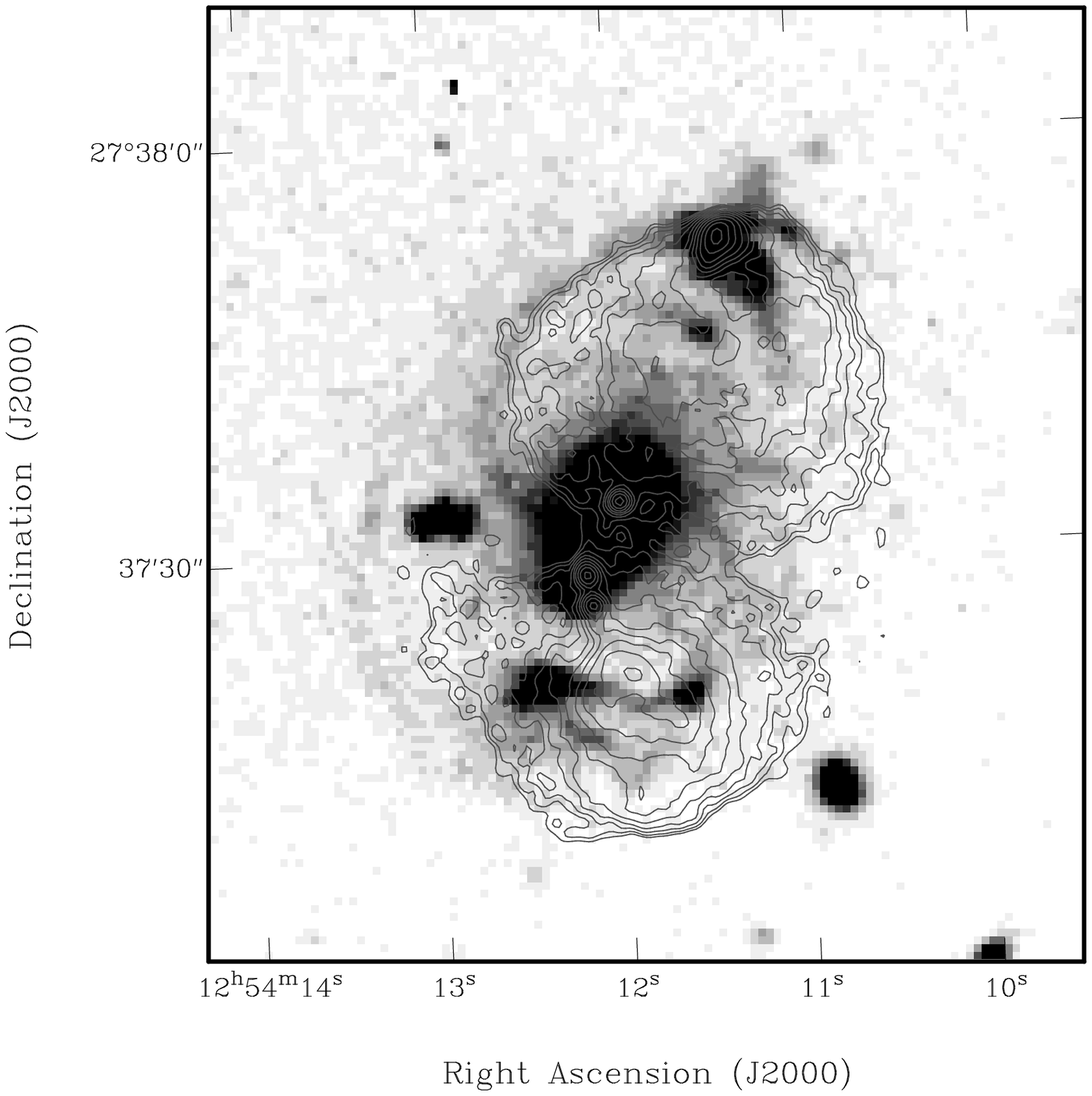,width=8cm}}
\caption{({\sl Left}) \Halpha\ image of Coma A (from Tadhunter et al.\ 2000)
  and ({\sl Right}) same image with contours of a 20-cm continuum image (about
  1 arcsec resolution, from van Breugel et al.\ 1985) superimposed.}
\end{figure*}

If mergers/accretions are indeed important in the evolution/formation of radio
galaxies, the question arises what the connection is between their gas
properties and those of other early-type galaxies, and whether we can learn
something from such a possible connection about the processes that are
involved in the AGN activity (Kauffmann \& Haehnelt 2000). Although a general
connection merger-AGN activity is reasonably well established, a number of
issues is still open, like, for example, 1) is the activity triggered by major
mergers between gas-rich galaxies or by minor accretion events, 2) what is the
relationship with other types of merging systems (e.g.\ the ultra-luminous
infra-red galaxies), 3) at what stage of the merger do the jets and the
associated activity occur, and 4) are all early-type galaxies likely to go
through phases of radio activity?

Powerful radio galaxies are also frequently associated with kpc-sized optical
emission line nebulosities, extending up to tens of kpc from the nucleus.  In
particular, the ionization of the gas is believed to be due to either
anisotropic UV radiation from the active nucleus or to shocks induced by the
radio jet and/or the radio lobes.  It is likely that in many cases both
mechanisms need to be invoked to explain the emission-line properties.  One
important element that may help to distinguish between the relative importance
of the two mechanisms would be to know the distribution of the {\sl neutral}
gas in and around the radio galaxies rather than only the ionization patterns
induced by jet or AGN.

It is clear from the above that neutral hydrogen studies can provide a good
tool to investigate these different aspects.  Usually the \HI\ is too weak to
be detected in emission in relatively far away radio galaxies, however certain
types of radio galaxies seem to be more promising than others for a possible
detection.  Among the good candidates is the radio galaxy Coma~A. There are
several observations that indicate that a complex interaction between the
radio structure and a rich interstellar (ISM) medium is occuring in this radio
galaxy (Sec.\ 2).  To obtain a more complete knowledge of the ISM in Coma~A, we
have observed this object in the 21-cm line of HI to try to detect, either in
emission or absorption, the gaseous environment of Coma~A.


\section{What is known about Coma~A}

Coma~A (3C277.3) is a well-known radio galaxy ($z = 0.08579$, Clark 1996)
studied in detail by e.g.\ van Breugel et al.\ (1985). The radio structure and
the radio power ($\log {\rm P} = 25.82$ W Hz$^{-1}$ at 1.4 GHz) is
characteristic of a radio galaxy intermediate between Fanaroff-Riley type I
(FRI) and II (FRII), although Coma A is often classified as FRII.  In Fig.~1
the radio contours from $\sim 1$ arcsec resolution VLA observations are shown
(see van Breugel et al.\ 1985 for more details).  The radio emission is mainly
formed by two ``fat'' lobes.  A bright hotspot is observed only in the
northern lobe.  The total size is $\sim 100$ kpc\footnote{Throughout this
  paper we use a Hubble constant of \( H_{\circ }=50 \) km s\( ^{-1} \) Mpc\(
  ^{-1} \), and $q_{\circ}=0$. At the redshift of Coma~A 1~arcsec corresponds
  to 2.2 kpc}.

HST observations (Capetti et al.\ 2000) show a prominent unresolved optical
core and patches of dust.  A jet-cloud interaction is occurring in Coma~A as
shown by the cloud of ionised gas observed next to a radio blob just south of
the nucleus.  This is clearly seen in HST images (Martel et al.\ 1999, Capetti
et al.\ 2000, Tadhunter et al.\ in prep). At this location, the southern radio
jet appears to be deflected, it ``lights up'' and finally it expands into the
southern lobe (van Breugel et al.\ 1985).

A spectacular system of interlocking emission-line arcs and
filaments of the same extent as the radio emission has been observed in
\Halpha\ (Tadhunter et al.\ 2000). There is a striking match between the radio
emission and the morphology of the ionised gas, in particular the optical
filament that curves around the northern radio lobe. This suggests that in
Coma A the direct interaction with the radio jet ionises the gas, but that
also the expansion of the lobe into the environment plays a role, and {\it
  vice versa} that the gaseous environment determines to some extent the
morphology of the radio continuum emission.

\begin{figure*}
\centerline{\psfig{figure=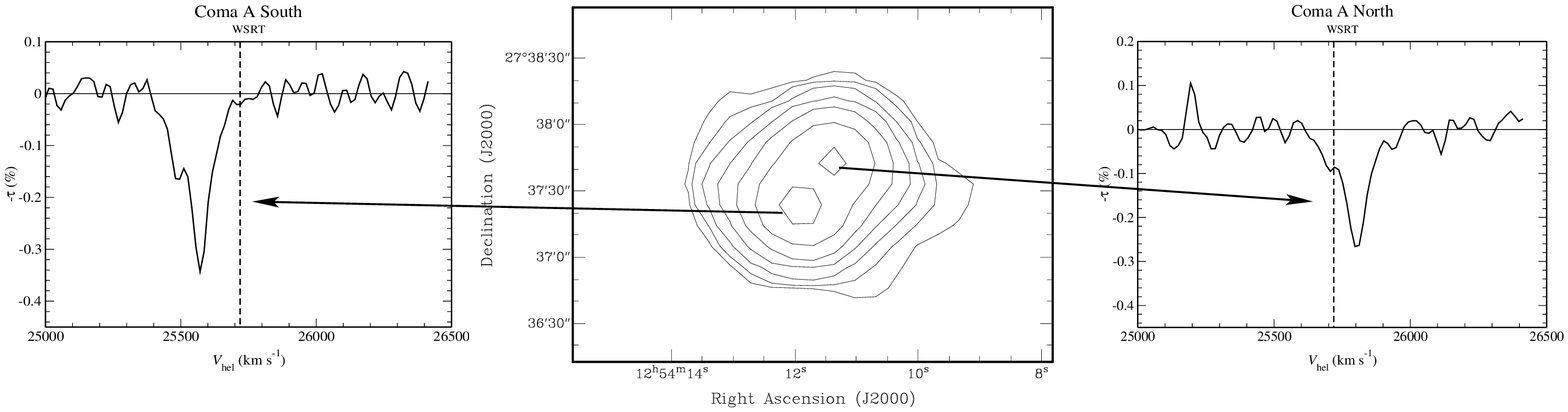,width=16cm}}
\caption{Profiles of the optical depth ($\tau$) of the 
\smallHI\ absorption against the two radio lobes as detected from the WSRT
  observations.  The contour levels are: 4 \mJybeam\ to 764 \mJybeam\ in steps
  of factor 2.4.}
\end{figure*}

\section{\HI\ absorption against the radio lobe}

\subsection {WSRT observations}

We have observed Coma~A at the frequency of the redshifted \HI\ line (1308.3
MHz) with the WSRT equipped with the DZB backend using a 10 MHz bandwidth, 128
channels, 2 polarisations and all 14 telescopes. The observations were done in
5 observing sessions, four of them of 12 h (14 Mar 2000, 20 March 2000, 19 Jul
2000) while from the last observation (03 Sep 2000) only 9 h of good data were
obtained due to technical problems.  This large amount of time is required in
order to reach good sensitivity and investigate the presence of \HI\ in
emission around the radio galaxy.

The data were calibrated and reduced using the Miriad package (Sault et al.\ 
1995) following standard reduction procedures, with the only exception that
particular care was taken in checking the stability of the bandpasses of the
antennae. This allowed us to reach a spectral dynamic range sufficiently high
(better than a few times $ 10^4$) to be limited by the thermal noise. The data
cube we choose to use was made using a robust-Briggs' weighting (Briggs 1995)
with robustness of zero.  The noise per channel is $\sim 0.22$ mJy/beam and
the beam size $38\times 15$ arcsec with p.a.\ $2^\circ$. The spectral
resolution is $\sim 18.7$ \kms. A continuum image was made using the line-free
channels.  This image was made using uniform weighting (equivalent to a robust
image with robustness = --2) in order to obtain a somewhat higher spatial
resolution ($22\times 11$ arcsec in p.a.\ $2^\circ$).

\subsection{The \HI\ absorption }
  
The continuum image of Coma A, as made from the WSRT data, is shown in Fig.\ 
2.  Given the relatively low spatial resolution, almost all the details
visible in Fig.\ 1 are smoothed out. The two radio lobes are just resolved,
while an extended component is also visible. The data clearly show that \HI\ 
absorption is detected against {\sl both} lobes at different velocities.  This
makes Coma A the first radio galaxy known where such absorption is seen at
large distances ($\sim$ 30 kpc) from the nucleus.  In the southern lobe, the
peak absorption is 4.9 \mJybeam\ centred at a velocity of $V_{\rm hel} =
25560$ \kms. The width of the absorption is FWHM $\sim 95$ \kms. In the
northern lobe, the peak absorption is 3.6 \mJybeam\ at a velocity of $V_{\rm
  hel} = 25780$ \kms\ and the width is FWHM $\sim 115$ \kms. The systemic
velocity of Coma A is $V_{\rm hel} = 25719$ \kms\ (Clark 1996).  Hence, the
absorption is blueshifted with respect to the systemic velocity in the
southern lobe and redshifted in the northern lobe.

The depth of an absorption line ($\Delta S$) depends on the optical depth
($\tau$), the continuum flux density ($S$) and the covering factor $c_{\rm f}$
as $\Delta S = c_{\rm f} S (1-e^\tau)$. Assuming for the moment a covering
factor $c_{\rm f} = 1$, the peak optical depths are $\tau = 0.35\%$ and $\tau
= 0.27\%$ for the southern and the northern lobe respectively.

The column densities of the neutral hydrogen follow from $N_{\matHI} = 1.83
\cdot 10^{18} T_{\rm spin} \int \tau dv$ where \tspin\ is the spin temperature
in Kelvin and $v$ is the velocity in \kms. Assuming the canonical $T_{\rm
  spin} = 100 $ K, we find column densities of $7.2 \cdot 10^{19}$ \atms\ for
the southern lobe and $6.1 \cdot 10^{19}$ \atms\ for the northern lobe.

\subsection{\HI\ emission}

No \HI\ was detected in emission in the immediate vicinity of Coma A. The
3-$\sigma$ upper limit to \HI\ emission in a single resolution element of the
cube is $4.3 \cdot 10^{19}$ \atms. Over the width of the absorption lines, the
3-$\sigma$ upper limit is $7$-$8 \cdot 10^{19}$ \atms, more or less the
column densities seen in absorption in the WSRT data. Thus, if there was
extended \HI\ emission associated with Coma A it would have been detected (the
VLA data show that the actual column densities are somewhat higher, see Sec.\ 
4). It is therefore likely that the extent of the neutral hydrogen associated
with Coma A is limited to the extent of the radio lobes and of the optical
emission line gas. 


None of the galaxies seen (in projection) near Coma A are detected in
emission. At a larger distance \HI\ is detected at the location of a small
galaxy at $\alpha = 12^{\rm h}54^{\rm m}41.8^{\rm s}$ $\delta = 27^\circ
30^\prime 15.2^{\prime\prime}$ (J2000) with $m_{R} =18.8 \pm 0.35$ (Odewahn et
al.\ 1995).  We derive a systemic velocity for this galaxy of $V_{\rm hel}
\sim 25900$ \kms\ and the \HI\ profile has a FWHM $\sim 300$ \kms.  This
galaxy is situated $\sim 15^\prime$ from Coma~A, corresponding to almost 2
Mpc. Thus, no interaction is expected between this galaxy and Coma~A. The \HI\ 
mass is substantial: $M_{\matHI} \sim 10^{10}$ \msun.

\begin{figure}
\centerline{\psfig{figure=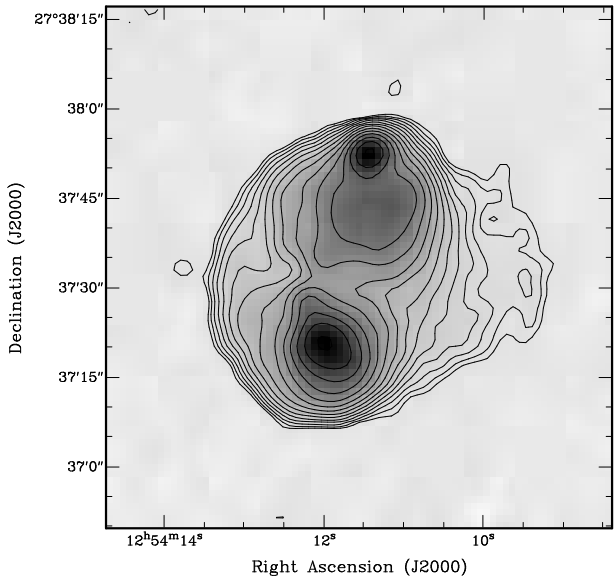,width=8.3cm,angle=0}}
\caption{Contours  
  of the radio continuum  at the resolution of $4.5 \times 4$
  arcsec (p.a.\ $-70^{\circ}$).  The contours for the continuum range
  from 1 mJy beam$^{-1}$ to 1 Jy beam$^{-1}$ in steps of factor
  1.5.}
\end{figure}

\section{The spatial distribution of the \HI\ absorption}

\subsection {VLA observations}

To be better able to locate where the \HI\ is occurring, we have observed
Coma~A also with the VLA in its B configuration.  These observations were
carried out on 15/16 May 2001.  In order to have sufficient velocity resolution
and to cover a large enough velocity range, we used a bandwidth of 3.125 MHz
and 64 channels for each of the two bands available.  The two bands were
placed with some overlap so that a total band of about 6~MHz is covered.  A
velocity resolution of 12 \kms\  is obtained using this setup.
The observation resulted in 6 hours of good quality data while the data of
four hours of observations could not be used due to technical problems.

Also the VLA data were reduced using the Miriad package, again using standard
reduction procedures.  Initially, two data cubes (one for each band) were made
using a robust-Briggs' weighting of zero (Briggs  1995).  This gives a
spatial resolution of $4.5 \times 4.0$ arcsec (p.a.\ $-70^\circ$).  The noise
in a single channel is 0.62 \mJybeam.  At this resolution, \HI\ absorption is
only weakly detected, especially in the northern lobe.  We therefore have
produced data cubes by tapering the visibilities in order to improve the
sensitivity for more diffuse absorption.  For the southern lobe, we have
tapered the data with a Gaussian of FWHM $ = 8$ arcsec, resulting in a beam of
$9.4 \times 9.0$ (p.a.\ 69$^\circ$) and a rms noise of 0.78 \mJybeam\ in a
single channel.  A Gaussian taper of FWHM $ = 10$ arcsec was necessary to image
the region of the \HI\ absorption in the northern lobe.  The resulting beam is
$11.5 \times 11.0$ (p.a.\ 66$^\circ$) and the rms noise is 0.85 \mJybeam\ per
channel.

A continuum image was also obtained using the line-free channels taken from
one of the observing bands.  The noise of the continuum image made with
robust-Briggs' weighting of zero is 0.37 \mJybeam.

\begin{figure}
\centerline{\psfig{figure=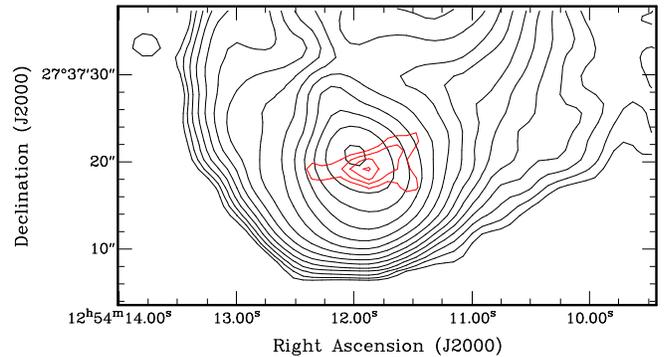,width=9cm,angle=0}}
\caption{Contours  
  of the radio continuum (black) and  contours of column density derived
  from the \smallHI\ absorption (red contours) obtained from the highest
  resolution data cube in the southern lobe (see text).
 The contours for the continuum range from 1 mJy beam$^{-1}$ to 1 Jy
 beam$^{-1}$ in steps of a factor 1.5. The contours of the  \smallHI\ absorption
  range from 0.4 to 1.0 $ \cdot 10^{20}$ \atms\ in step of $0.2 \cdot 10^{20}$
  \atms.}
\end{figure}

\begin{figure}
\centerline{\psfig{figure=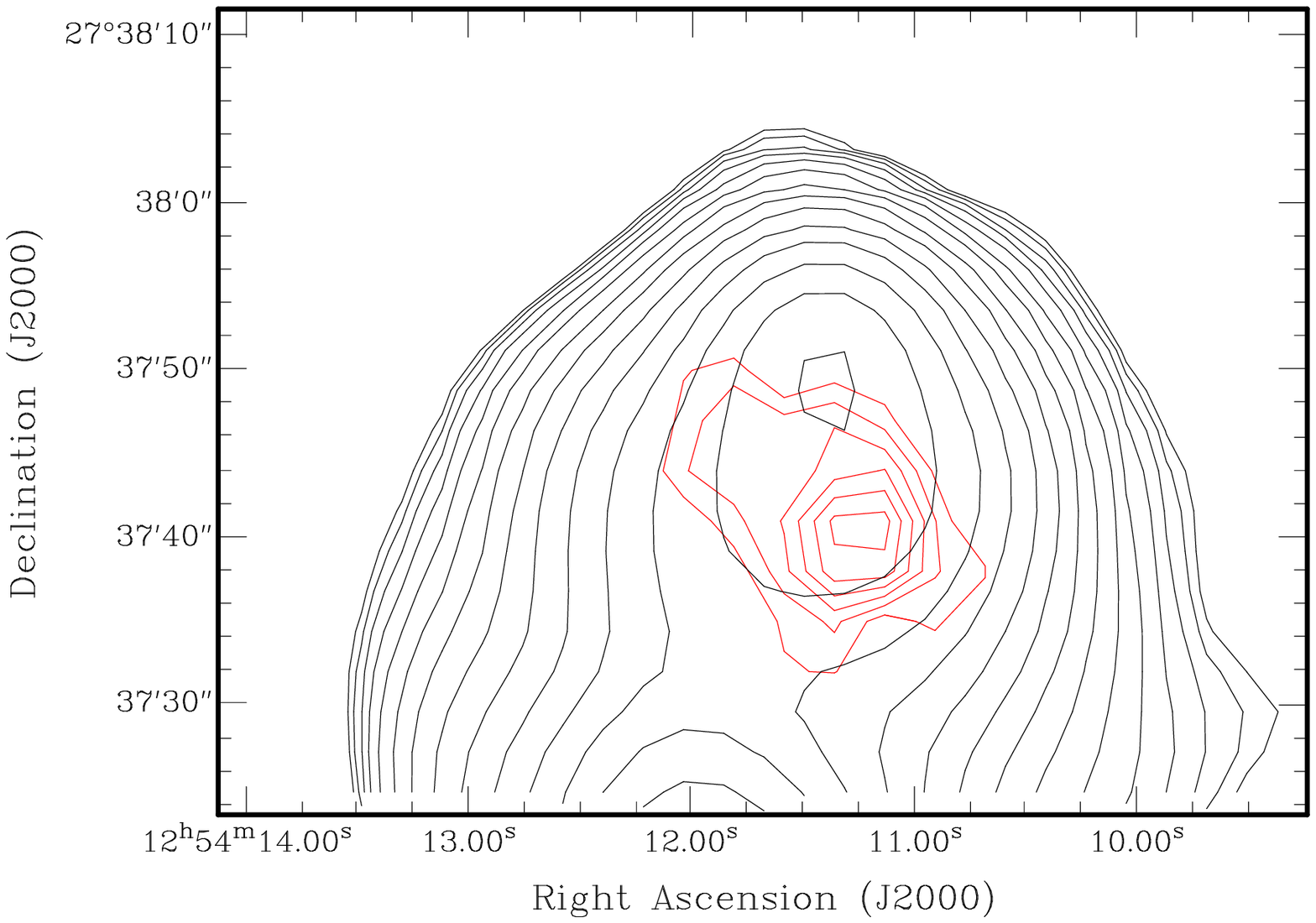,width=9cm,angle=0}}
\caption{Contours  
  of the radio continuum (black) and  contours of column density derived
  from the \smallHI\ absorption (red contours) obtained from the highest
  resolution data cube in the southern lobe (see text).
 The contours for the continuum range from 2 mJy beam$^{-1}$ to 1 Jy
 beam$^{-1}$ in step of factor 1.5. The contours of the  \smallHI\ absorption
  range from 0.4 to 0.9 $ \cdot 10^{20}$ \atms\ in step of $0.1 \cdot 10^{20}$
  \atms.}
\end{figure}

\begin{figure*}
\centerline{\psfig{figure=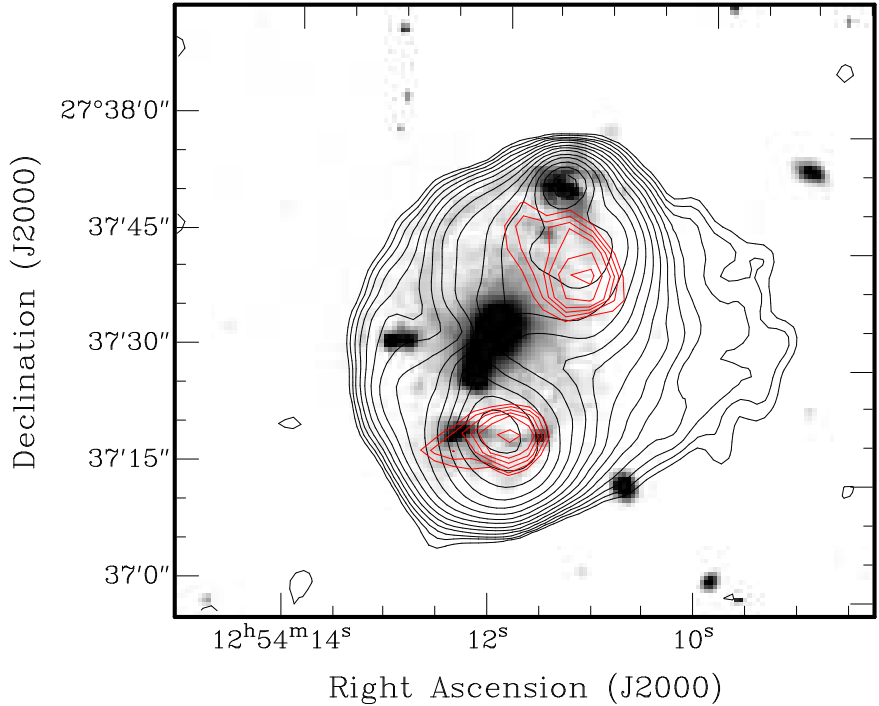,width=12cm,angle=0}}
\caption{\Halpha\ image (grey scale) with superimposed the contours  
  of the radio continuum (grey) and the contours of column density derived
  from the \smallHI\ absorption (black). In the northern  lobe the contours
  range from 0.35 to 0.85 $ \cdot 10^{20}$ \atms\ in step of $0.1 \cdot 10^{20}$
  \atms. In the southern lobe the contours range from 1 to 1.5 $ \cdot
  10^{20}$ \atms\ in step of $0.1 \cdot 10^{20}$ \atms.}
\end{figure*}

\subsection{The location  of the \HI\ absorption}

In Fig.\ 3 the continuum image as made from the VLA data is
shown. Much more detail is visible compared to the image made from the
WSRT data. The two radio lobes are now well resolved, while still much
of the extended component that is visible in the WSRT image, but not
in the image made with the VLA A array, is detected.

In the full resolution VLA data cubes very little absorption is
reliably detected. No absorption is detected against the nucleus nor
against the location just south of the nucleus where the jet-cloud
interaction is occurring (van Bruegel et al.\ 1985), with upper limits
to the optical depth of 4 and 2.7 \% respectively.

In the southern lobe the peak absorption is detected only at the 3-$\sigma$
level at 2.9 \mJybeam\ in a single channel. This corresponds to a maximum
optical depth of $\tau = 1.7\%$.    In Fig.~4 the integrated \HI\ 
  absorption is shown (red contours) at the full resolution. The absorption is
  found near the region with the brightest continuum emission, with a
  suggestion that the absorption occurring just SW of this maximum in the
  continuum.  No absorption is detected above the 3-$\sigma$ level in the
  northern lobe.  These results indicate that the \HI\ absorption is due to an
  {\sl extended} screen in front of the radio lobes.

In order to improve the sensitivity for extended absorption lower resolution
data cubes were made as described in Sec.\ 4.1. In these lower-resolution
cubes absorption is detected against both radio lobes.   Figure 5 shows
  where this absorption is detected in the northern radio lobe (at the
  resolution of the tapered data). The red contours represent the integrated
  \HI\ absorption.  The absorption is somewhat extended and is not
occurring near the continuum hotspot.  It is clearly displaced from it,
in the direction of the nucleus.

 In Fig~6 the contours of the \HI\ absorption detected against both radio
lobes (at the resolution of the tapered data) are shown, superimposed on
the \Halpha\ image.
 
The peak absorption in the southern lobe is 6.2 mJy beam$^{-1}$ which
corresponds to an optical depth of $\tau_{\rm max} = 1$ \%. This is higher
than in the WSRT data, but this can be explained by the difference in velocity
resolution.  The width of the absorption profile is FWHM $\sim 70$ \kms.  For
the northern lobe the peak of the absorption is 4.2 mJy beam$^-1$ ($\tau_{\rm
  max} = 0.9$ \%) and FWHM $\sim 60$ \kms.  The corresponding column densities
are $1.3 \cdot 10^{20}$ \atms\ for the southern lobe and $8.7 \cdot 10^{19}$
\atms\ for the northern lobe, assuming $T_{\rm spin} =100$ K.  By integrating
over the area where the absorption is occurring, the estimated \HI\ mass is
found to be about $M_{\matHI} \simeq 10^9$ \msun.

When comparing with the WSRT results one sees that the values of the peak
absorption are, within the errors, the same, but that the widths of the
absorption profiles are smaller in the VLA data (see Fig.\ 7). The VLA
observations recover about 70-75\% of the absorption detected in the WSRT
data. This, and the larger widths of the WSRT profiles implies that more
absorption, at a level too faint to be detected in the VLA observations, is
present over a larger region of the lobes, at somewhat different velocities.
The absorption therefore covers a significant fraction of both radio lobes.

\begin{figure}
\centerline{\psfig{figure=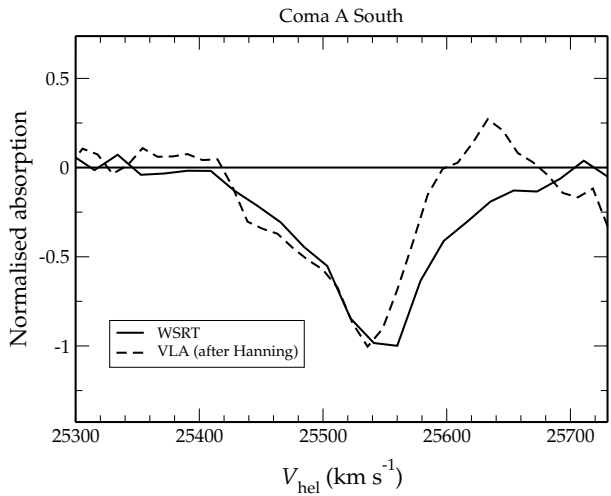,width=8cm,angle=0}}
\caption{Comparison between the profile of the south lobe obtained with
WSRT (long dash) and the VLA after tapering (short dash).  
 }
\end{figure}

\section{Neutral and ionised gas: A possible large-scale gas disk}

The data discussed above show that there is \HI\ absorption in front of both
radio lobes of Coma~A.  This makes Coma A quite unique: the
absorption is not detected against the nuclear regions of a radio galaxy but
it is situated at large distances (tens of kpc) from the centre.

The absorption covers a significant fraction of both radio lobes. The
strongest absorption is localised in areas at about 15 arcsec ($\sim 33$ kpc)
from the radio core.  We do not detect \HI\ absorption against some of the
brighter regions of radio emission, e.g.\ the hot spot in the northern lobe
and the bright knot just south of the nucleus.

It is important to understand the relationship between the neutral hydrogen
detected at large distances from the centre and the large structure of ionised
gas found by Tadhunter et al.\ (2000). For this it is interesting to compare the
velocities of the \HI\ with those of the ionised gas (from Clark 1996). 
These optical data consist of long-slit spectra taken along the major axis of
the ionised-gas distribution.   These data do not cover exactly the same
regions where the \HI\ absorption is detected, but the difference in location
is not too large and a comparison is still meaningful. 

The match between the optical velocities and those of the \HI\ is very good
(Fig.\ 8). Both the ionised gas and the neutral gas are blueshifted, by about
the same amount, south of the centre and redshifted north of the centre. This
close correspondence in velocity between the ionised and neutral gas
demonstrates that they are two phases of the same gas structure.

From the morphology of the ionised gas, Tadhunter et al.\ (2000) argued that
Coma~A could have experienced a recent merger and the filamentary structure of
the ionised gas would represent pre-existing gas structures.  The large amount
of neutral gas detected in this galaxy via \HI\ absorption, supports the idea
that Coma~A is likely to have experienced a merger. Considering that a large
fraction of the total gas mass must be ionised, this merger must have involved
at least several times $10^9$ \msun\ of gas. The merger in Coma A must have
involved at least a reasonably sized gas-rich galaxy.

A number of faint galaxies are found within 100~kpc
of the nucleus of Coma~A and could be  associated with the
extended \Halpha\ filamentary structures. One of them (see Fig.1 in Tadhunter
et al.\ 2000) is situated (in projection) on the western edge of the southern
\HI\ cloud.
This galaxy, even if spectroscopically confirmed to be a real companion,
appears too small to be the donor of the observed gas because of the large
amount of gas detected around Coma~A and the large extent over which this gas
is found.

\begin{figure}
\centerline{\psfig{figure=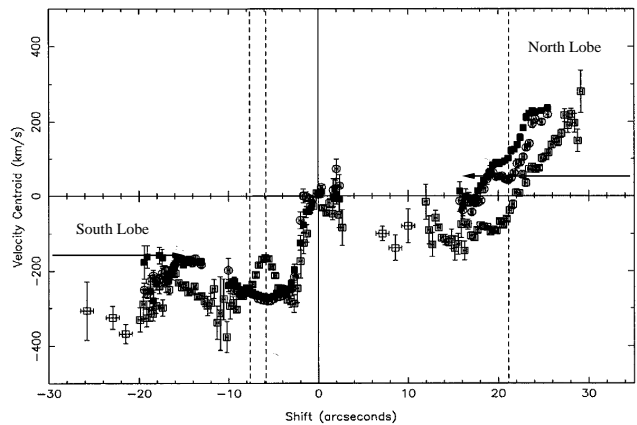,width=9cm,angle=-90}}
\caption{ Plot of the optical velocities along the radio axis derived from
  \Halpha\ (filled squares) and [\smallOIII] (open squares) emission lines.
  The velocities derived from the \smallHI\ absorption against the two radio
  lobes are marked.}
\end{figure}

One of the aims of the observations presented here was to look for very
large-scale \HI\ tails around Coma~A.  These structures are often observed in
recent mergers (e.g.\ Yun et al.\ 1994; Hibbard \& van Gorkom 1996; Hibbard \&
Mihos 1995).  No such structures have been detected in Coma~A.  The gas
acquired  is apparently reasonably settled in an area of
a few tens of kpc, instead of the hundred kpc long structures often seen in
on-going mergers.  Moreover, the kinematics  and the more or less
symmetric distribution of the gas, suggest that the gas has settled at least
to some extent. Considering that this settling, given the size and the
velocities observed, would take roughly a few times $10^8$ yrs, the merger in
Coma A is likely to have happened several times $10^8$ yrs ago.

An important constraint to consider in the interpretation is that \HI\ 
absorption is detected against {\sl both} radio lobes.  Neutral gas must be,
therefore, present in front of both of them.  Thus, to explain the observed
characteristics of both the ionised gas and the neutral hydrogen absorption in
Coma~A, we have to assume that if the gas is distributed in a disk-like
structure, the radio plasma is actually expanding {\sl in} this disk. The
distribution of the ionised gas supports this.  The morphology of the ionised
gas appears to be distorted by the interaction with the radio lobes, e.g.\ 
with shocks fronts sweeping up material into the filamentary structures
observed.  The fact that the neutral hydrogen seems to show quite broad
absorption profiles could be an indication of a kinematical disturbance as the
effect of interaction between the radio and the ISM. A first-order
anti-correlation is observed between the ionised and the neutral gas, in
particular in the northern lobe.  

It is also interesting to note that we find the \HI\ in the region where we
would expect the gas to be ionised if there is an illuminating quasar.
Therefore, the \HI\ measurements provide further evidence for the idea that
the quasar illumination in Coma~A is weak or absent (Tadhunter et al.\ 2000).

In many cases where a gas/dust disk is observed in a radio galaxy, the radio
axis is perpendicular to this disk.  However, this is not a golden rule and
applies mainly to small dust lanes near the nucleus (e.g.\ de Koff et al.\ 2000,
de Ruiter et al.\ 2001).  Moreover, several examples exist, perhaps related to
the tri-axial nature of the mass distribution, where the inner gas disk is
perpendicular to the outer gas disk (e.g.\ NGC 5266, Morganti et al.\ 1997;
NGC 2685, Oosterloo et al.\ (in prep.) and possibly also Centaurus A,
Schiminovich et al.\  1994).  In some of these galaxies the radio jet is
directed into the plane of the outer gas disk and interacting with this gas
disk. A spectacular example of this is the Seyfert galaxy IC 5063 (Morganti et
al.\ 1998). Hence, a geometry with the lobes expanding into the outer gas disk
can be expected and is actually observed in some cases.

Note that in the radio continuum, apart from the two lobes, there are
extensions to the east and the west with the western extension larger than the
eastern one. The morphology of the ionised gas suggests that the interaction
on the eastern side is stronger than on the western side. The larger western
extension of the radio continuum on suggests that the gas disk plays a role in
confining and shaping the radio continuum structure.

\section {How unique is Coma~A?}

Large disks of neutral gas have already been observed  in a number of
early-type galaxies (see e.g.\ Oosterloo et al.\ 2001a,b and ref. therein; van
Gorkom 1997 and ref. therein) and the detection of such a disk in Coma~A
fits naturally in the scenario that  these large disks and the AGN
activity are merger related. The main difference of Coma~A with the galaxies
mentioned above is the higher radio continuum power.  Such extended gas disks
are very difficult to detect if the galaxy is at a relatively large redshift
(as most of the radio galaxies are), but our work on Coma~A shows that such
disks can be studied through \HI\ absorption, an observational technique that
can, with current radio telescopes, be used out to redshift $\sim$3.5.

The only other objects known to us that appears to show characteristics of the
neutral gas distribution similar to Coma~A are 3C~433 and 3C~234.  Recent VLA
observations (Morganti et al.\ 2001) show that at least part of the \HI\ 
absorption (detected originally with the Arecibo telescope, Mirabel 1989) is
situated against the southern radio lobe at about 40 kpc from the nucleus.
Unlike Coma~A, no ionised gas has been detected in 3C~433 at such a distance
from the centre.  In 3C~234 \HI\ absorption has been detected toward the
  NE lobe at about 50 arcsec from the centre
  (Pihlstr\"om, 2001).

On a smaller scale, situations similar to Coma~A can be found in a
handful of objects. We want to stress  that in these cases the scale of the
phenomenon is at most a few kpc while in Coma~A the absorption is observed at
tens of kpc from the nucleus.

One of these objects is  PKS~2322--123, the central radio galaxy in the 
cooling flow cluster Abell 2597 studied by O'Dea et al.\ (1994) and Taylor et
al.\ (1999).  Spatially extended 21-cm \HI\ absorption is detected against
this source (O'Dea et al.\ 1994) indicating that the radio lobes are surrounded
by a collection of clouds containing both neutral and ionised gas.
Interestingly  this object also shows evidence of optical arcs associated with
the radio lobes, as observed in the case of Coma~A. Koekemoer et al.\ (1999)
suggest that the properties of this source are explained in terms of accretion
of gas by the cD during the interaction with a gas-rich galaxy.  The same
could apply to Coma A, with the main difference that in PKS~2322--123 it
occurs  at only a few kpc from the nucleus.

Two more examples where the \HI\ absorption is observed in coincidence with
the radio lobes instead of the radio core are IC~5063 (Morganti et al.\ 1998,
Oosterloo et al.\ 2000) and 3C~236 (Conway \& Schilizzi 2000).  Again, in both
cases the distance from the nucleus of the \HI\ absorption is only a few kpc.
These are  examples of a jet-cloud interaction between the radio
plasma and the ISM.  It is interesting to note that in IC~5063 (and likely
also in 3C~236) the radio axis is parallel to the axis of the large-scale gas
disk, as we claim could be the case in Coma~A.

More radio galaxies with very extended neutral gas distributions are likely to
exist.  Extended \HI\ absorption (observed against the Ly$\alpha$
  emission) has been found in a high fraction of high-$z$ radio galaxies (van
  Ojik et al.\ 1997).  This is considered an indication that high-$z$ radio
  galaxies are located in dense environments and is a diagnostic for
  the effects of radio jet propagation in this dense medium.  Although this
  phenomenon may be occurring more frequently and more efficiently at high
  redshifts, in the low redshift radio galaxies described here we may
  witnessing a similar situation.  As the sensitivity of the radio telescopes
improves, such extended gas distributions will be found in more radio
galaxies. We have recently detected a large disk-like \HI\ structure in the
compact radio galaxy B2~0648+27 (Morganti et al.\ 2001). In this object the
radio source is unresolved with arcsecond resolution and therefore \HI\ 
absorption is only detected against the central source.  However, \HI\ in
emission is detected out to a radius of $\sim 80$ kpc.

\section{Summary}

We have detected \HI\ absorption against the radio lobes of the
radio galaxy Coma A. This absorption covers a significant fraction of these
lobes and occurs at large distances from the nucleus ($\sim 33$ kpc). The
kinematics of the \HI\ matches that of the large structure of ionised gas
present in Coma A and the ionised and the neutral gas are two phases of the
same structure. No \HI\ is detected in emission outside the area covered by
the lobes. The fact that we detect absorption on {\sl both} sides from the
nucleus suggests that the radio  lobes are expanding into a large gas
disk. The morphology of the ionised gas 
strongly supports this. 

The gas disk appears relatively settled indicating that the AGN activity,
assuming the period of such activity lasts at most a few times $10^7$ yrs,
started at least $10^8$ yrs after the merger.  The mass of the gas involved in
this merger is at least $10^9$ \msun, suggesting that at least one
reasonably-sized gas-rich galaxy was involved.

\begin{acknowledgements}
The WSRT  is operated by the ASTRON (Netherlands Foundation for Research in
Astronomy) with support from the Netherlands Foundation for Scientific Research
(NWO). The VLA is a facility of the National Radio Astronomy Observatory,
which is operated by Associated Universities Inc., under cooperative agreement
with the National Science Foundation.
\end{acknowledgements}


\end{document}